\documentstyle[tighten,preprint,prb,aps,psfig]{revtex} 
\begin{document} 
\draft 
\title{Spatial structure of an incompressible Quantum Hall strip}
\author{Ivan A. Larkin\dag and L.~S.~Levitov\ddag} 
\address{\dag 
Department of Physics and Astronomy,
University of Sheffield, Sheffield  S3 7RH, UK\\ 
\ddag 
Center for Materials Science \& Engineering, Physics Department,
MIT, Cambridge, MA-02139}
\maketitle 
 
\centerline{\bf Abstract} 
 
The incompressible Quantum Hall strip is sensitive to charging of localized 
states in the cyclotron gap. We study the effect of localized states
by a density functional approach and find electron density and the strip 
width as a function of the density of states in the gap. 
Another important effect is electron exchange. By using a model density 
functional which accounts for negative compressibility of the QH state, 
we find electron density around the strip. At large exchange, the density 
profile becomes nonmonotonic, indicating formation of a 1D Wigner crystal at
the strip edge. Both effects, localized states and exchange, lead to 
a substantial increase of the strip width.
\\

\centerline{\bf 1. Introduction} 

\noindent
Theory of the QHE predicts that near integer filling the system 
divides into compressible regions separated by incompressible 
strips [1]. Potential distribution within a QHE sample was 
recently imaged by using capacitance probes [2,3], atomic force 
microscope [4] and single electron transistor [5]. High 
resolution images of the incompressible strip [2,3] give the 
strip width several times larger than the theoretical prediction 
[1]. To bridge between theory and experiment one has to extend 
the analysis [1] to include\\
  $\bullet$ the effect of disorder 
producing finite density of states in the cyclotron gap;\\
  $\bullet$ electron exchange correlations which affect compressibility 
of the QHE state.\\
  Below we study these effects using a Density
Functional approach, taking special care of the
effect of a large dielectric 
constant ($\epsilon_{\rm GaAs}=12.1$). Because of relatively small 
depth of the 2DEG beneath the semiconductor surface, the 
interparticle interaction within the 2DEG is affected by image 
charges. This changes electrostatics of the strip, and modifies 
potential induced on the exposed surface. 
 
Finite density of states in the QHE gap gives rise to 
a finite screening length. For an incompressible strip of width 
exceeding this screening length, we find a large departure from 
the theory [1], in agreement with [8]. The results
compare well with the experiment [3].
 
The effect of electron exchange is important in determining the 
structure of compressible regions 
adjacent to the strip. Exchange correlation gives rise to negative 
compressibility [7] of the 2DEG. We consider negative compressibility
by using a model density functional, and show that it 
strongly alters the distribution of electric 
charge, even to the extent that the potential and the charge 
density profiles can become nonmonotonic. 
 
\centerline{\bf 2. The effect of finite density of states in the cyclotron gap} 
 
\noindent
Incompressible strips are formed in the regions of nonuniform 2DEG
density, at nearly integer filling, created either by perturbing 
the exposed surface by an STM probe [2] or by gating the 2DEG [3]. 
The strips are aligned normal to the average 
2DEG density gradient. Charge distribution around 
the strip is controlled by electrostatics [1,6].
 
Density $n(r)$ in the 2DEG buried at a distance $d$ beneath 
semiconductor surface can be found by minimizing a density functional: 
  \begin{equation}\label{DF} 
-U_{\rm ext}(r)=\int V(r-r')n(r')d^2r' +\mu(n)\ ,\qquad 
V(r)=\frac{e^2}{\epsilon |r|}+\frac{(\epsilon-1)e^2}{\epsilon 
(\epsilon+1)\sqrt{r^2+4d^2}} 
  \end{equation} 
Here $U_{\rm ext}(r)$ is external potential due to donors or 
gates, the Hartree interaction $V(r-r')$ takes into account 
image charges, and the chemical potential $\mu$ includes various
non--Hartree contributions: 
finite density of states, QHE gap, exchange effects, etc.
 
Since the scale of the observed structures [2,3] is always 
larger than the 2DEG depth $d$, we replace the Hartree 
interaction in (\ref{DF}) by 
$V(r)=2e^2/((\epsilon+1)|r|)$, assuming $|r|\gg d$. 
 
In this section we consider a free electron model for $\mu(n)$ 
which includes degenerate ({\it i.e.}, infinitely compressible) 
Landau level states and localized states in the QH gap: 
  \begin{equation}\label{DOS} 
dn/d\mu= 
(n_{\rm LL}-n_{\rm gap}) 
 \sum_{m> 0}\delta(\mu-m \hbar\omega_c ) \ +\ 
n_{\rm gap}/\hbar\omega_c\ ,
\qquad n_{\rm LL}=eB/hc\ . 
  \end{equation} 
In the simplest model Eq.(\ref{DOS}) the density of localized 
states is constant. 
 
Below we focus on the $m=1$ QHE plateau. 
To introduce the 2DEG density gradient into the problem, we express $U_{\rm ext}$ 
in terms of fictitious positive charge density within the 2DEG plane: 
  \begin{equation} 
U_{\rm ext}(r)=\int V(r-r')n_{\rm eff}(r')d^2r' 
\ ,\qquad 
n_{\rm eff}(r)=-n_{\rm LL}-\vec r\cdot\vec\nabla n 
  \end{equation} 
For $\mu(n)\to0$ in (\ref{DF}), {\it i.e.} without 
magnetic field, it follows from (\ref{DF}) that
$n(r)=-n_{\rm eff}(r)$. 
 
Now, we nondimensionalize the problem by choosing
  \begin{equation}\label{w0n0}
w_0 =\left((\epsilon+1) \hbar\omega_c /2e^2|\nabla n|\right)^{1/2}
\qquad {\rm and}\qquad 
n_0=\left((\epsilon+1) \hbar\omega_c|\nabla n|/2e^2\right)^{1/2}
  \end{equation}
as the length and density units. 
Then the only remaining dimensionless parameter is the ratio 
of the fully incompressible strip width $w_0$ to the screening radius
$r_s$ corresponding to the density of states $n_{\rm gap}/\hbar\omega_c$ in the
gap. Up to a factor of $2\pi$, this ratio is
$\gamma=n_{\rm gap}/n_0=n_{\rm gap} \left(2e^2/(\epsilon+1) \hbar\omega_c|\nabla n|\right)^{1/2}$. 
The nondimensionalized problem reads: 
  \begin{equation} 
\int\frac{(x'-\delta n(r'))d^2r'}{|r-r'|}= 
\int\limits^{\delta n(r)}_0 F_\gamma (u)du 
  \end{equation} 
where $\delta n(r)=n(r)-n_{\rm LL}$, and 
$F_\gamma (u)=\gamma^{-1}$ for $|u|<\gamma/2$, and $0$ otherwise. Here the 
coordinate system is such that the $x$ axis is normal to the strip, and the
$y$ axis is parallel to the strip.

One can obtain exact results for $\gamma\to0$ and $\gamma\gg1$. The strip 
width at $\gamma=0$ is $2w_0/\pi$, in accord with the electrostatic
sproblem [1].
At $\gamma\gg1$ the deviation from constant density gradient is small,
because electrostatic potential is well screened. 
In this case, spatial variation of the chemical potential follows that 
of the density, increasing by $\hbar\omega_c$ across the strip. 
Hence the strip width is $n_{\rm gap}/|\nabla n|$. 

\begin{figure}
\centerline{\psfig{file=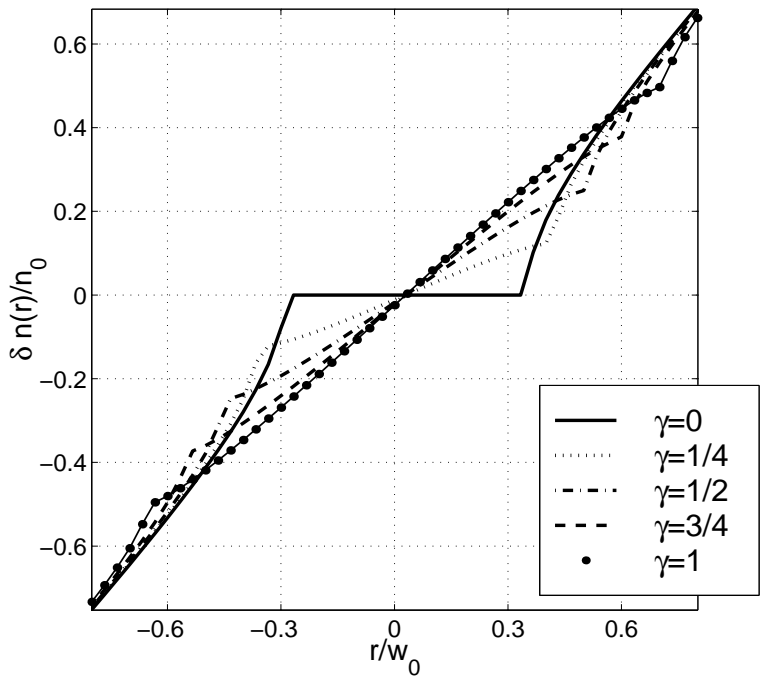,width=4.0in}}
\vspace{0.5cm}
\caption{The effect of localized states on $n(r)$
for 5 values of $\gamma=n_{\rm gap}/n_0$. The units used are
$w_0=\left((\epsilon+1) \hbar\omega_c /2e^2|\nabla n|\right)^{1/2}$ and 
$n_0=\left((\epsilon+1) \hbar\omega_c|\nabla n|/2e^2\right)^{1/2}$.}
\label{fig1}
\end{figure}

We solve the problem numerically for all $\gamma$ (see Fig.\ref{fig1}). 
In the whole range of $\gamma$ the strip width 
is reasonably accurately given by the formula 
  \begin{equation}\label{approx_width}
w\approx (2/\pi+\gamma)w_0=(2/\pi)w_0+n_{\rm gap}/|\nabla n| \ ,
  \end{equation}
interpolating between the two exactly solvable limits.

A common model for the density of states in a Landau level is a broad line
(gaussian or lorentzian) with the states localized in the tail. In this model,
the transition between (less compressible) strip and (more compressible)
outer region will be more gradual than in the model considered. The estimate (\ref{approx_width})
for the width of the strip, however, will remain correct, assuming that $n_{\rm gap}$ measures
total number of states in the Landau levels tails.

In the experiment [3], at the $m=2$ plateau, the density of states in the gap
$n_{\rm gap}\approx 0.03\, n_{\rm total}$, 
where $n_{\rm total}=1.5\cdot  10^{11}\ {\rm cm}^{-2}$.
The density gradient 
$\nabla n\approx 2\cdot 10^{10}\ {\rm cm}^{-2}/\mu{\rm m}$. 
Substituting this in (\ref{w0n0}),
we get $w_0=0.3\mu{\rm m}$, $\gamma\approx 1$. 
In the fully incompressible case [1],
the strip width would be $2w_0/\pi=0.2\,\mu{\rm m}$. 
The observed width $0.5\,\mu{\rm m}$ agrees 
with Eq.(\ref{approx_width}) for estimated $\gamma$. 

\centerline{\bf 3. The effect of negative compressibility of the compressible edge} 

\noindent
For a fully incompressible strip ($n_{\rm gap}=0$), the density 
is constant within it and varies outside as a square root of the 
distance from the strip edge [1]. Here we study how this 
behavior is modified due to finite compressibility of the Landau 
level states. The Thomas--Fermi theory recipe is to use 
Eq.(\ref{DF}) with $\mu(n)=\delta n/\kappa$, where $\kappa$ is 
compressibility. Such a model, however, is inconsistent, because 
of the {\it negative} sign of $\kappa$ in the QH state 
[7]. The Thomas--Fermi problem with $\kappa<0$ leads to 
an unphysical instability. 
 
The difficulty is circumvented by realizing that the exchange 
interaction in the case of negative compressibility is 
essentially nonlocal [9]. This motivates using in 
(\ref{DF}) an effective interaction which is simplest to write 
in the Fourier representation: 
  \begin{equation}\label{Veff} 
V_{\rm eff}(k)=\frac{4\pi e^2}{(\epsilon+1)|k|}\Lambda(k)\ ,\qquad 
\Lambda(k)>0,\qquad 
\Lambda(0)=1,\qquad 
\Lambda'(0)=\frac{(\epsilon+1)}{4\pi e^2\kappa}=-a\ ,
  \end{equation} 
where $a>0$ is the screening length. The interaction (\ref{Veff}) with listed 
restrictions on $\Lambda(k)$ ensures stability as well as 
correct Hartree interaction and compressibility. Otherwise, one 
can make a reasonable choice of $\Lambda(k)$ at $ka>1$. 
 
The problem (\ref{DF}) near the strip edge, with $V_{\rm eff}$ of the form 
(\ref{Veff}) accounting for exchange effects, can be solved by 
the Wiener--Hopf method [9]. 
For that, we write $\delta n(x)=n^{+}(x)\theta (x)+n^{-}(x)\theta (-x)$, 
where $x>0$ is the compressible region, and Fourier transform 
Eq.(\ref{DF}): 
   $V_{\rm eff}(k)n_k^{+}=-(U_{\rm ext}^{+}(k)+U_{\rm  ext}^{-}(k))$, 
where $n_k^{\pm}$ and $U_{\rm ext}^{\pm}(k)$ are analytic in the 
upper and lower complex $k$ half-planes. The Wiener--Hopf trick 
requires factoring $V_{\rm eff}(k)=A_k^{+}/A_k^{-}$, where $\pm$ 
indicates the analyticity half-plane. Then, 
$A_k^{-}U_{\rm ext}^{+}(k)=\left[ A_k^{-}U_{\rm ext}^{+}(k)\right]^{+} 
+\left[ A_k^{-}U_{\rm ext}^{+}(k)\right]^{-}$ which yields 
$n_k^{+}=-[A_k^{-}U_{\rm ext}^{+}(k)]^{+}/A_k^{+}$. 
 
We use $V_{\rm eff}(k)$ of the form (\ref{Veff}) with
  \begin{equation}\label{ourLambda}
\Lambda(k)= \exp\left[-ak(1-(2/\pi)\tan ^{-1}(k/\lambda))\right]\ ,
  \end{equation}
and obtain a Wiener--Hopf solution in a closed form. 
Here the parameter $\lambda$ regularizes the interaction at
large $k$ (and small $r$): 
$V_{\rm eff}(r\ll\lambda^{-1})=e^{-2\lambda a/\pi}V(r)$. 
Factoring this $V_{\rm eff}(k)$ gives 
  \begin{equation} 
A_k^{+}=\frac{4\pi e^2}{(k-i\delta )^{1/2}}\left( \frac{\delta +ik}{\lambda 
+ik} \right) ^{iak/\pi} 
\ ,\ \ 
A_k^{-}=(k+i\delta )^{1/2}\left( 
\frac{\delta -ik}{\lambda -ik}\right) ^{iak/\pi}\ , 
  \end{equation} 
where $\delta=+0$. Near the edge $U_{\rm ext}(x)=Ex+c$, and thus 
  \begin{equation} 
n_k^{+}=\frac{(\epsilon+1)E}{4\pi e^2}
\frac{(i\delta^\ast)^{1/2}}{(k-i\delta )^{3/2}}\left( \frac{k-i\lambda}{k -i\delta}\right)^{iak/\pi}\ , 
  \end{equation} 
where $\delta^\ast\sim w_0^{-1}$. The inverse Fourier transform of $n_k^{+}$ gives the charge 
distribution near the strip edge. Note the asymptotic behavior 
of $\delta n(x)$: $\delta n(x\!\gg\!\lambda^{-1})=2n_0(x/\pi w_0)^{1/2}$,\
$\delta n(x\!\ll\!\lambda^{-1})=2n_0(x/\pi w_0)^{1/2} e^{\lambda a/\pi}$. Here we expressed
$E$ and $\delta^\ast$ in terms of $w_0$ and $n_0$.

The solution shows that at large screening length $a$ there is 
a significant departure of the density near the edge
from the square root profile of [1]. The density profile becomes
{\it nonmonotonic} at $a\lambda\gg1$. The interpretation 
of this result is that as the density is lowered, 
electrons at the edge form
a one dimensional Wigner solid at densities such that
the interior of the system is still fluid. 

\begin{figure}
\centerline{\psfig{file=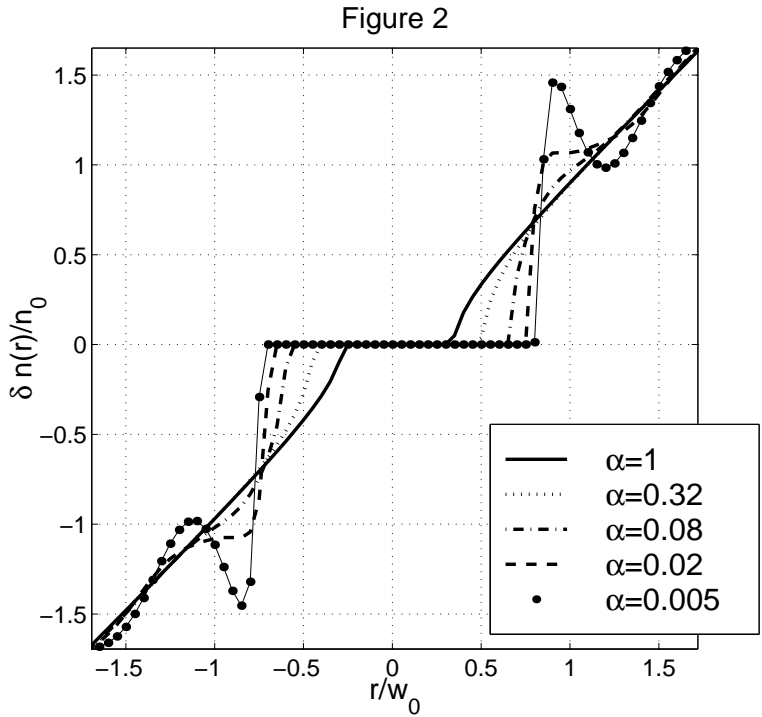,width=4.0in}}
\vspace{0.5cm}
\caption{ Incompressible strip density distribution created by effective
interaction (\protect{\ref{modelV}}) shown
for 5 values of the exchange parameter $\alpha$, with $\tilde a=0.5 w_0$.}
\label{fig2}
\end{figure}

We studied numerically the effect of exchange on the strip
width. As the exchange interaction parameter increases, the 
strip becomes wider (see Fig.\ref{fig2}). 
In the simulation, a model interaction
  \begin{equation}
  \label{modelV}
V_{\rm eff}(r)=\alpha/|r|+(1-\alpha)/(r^2+\tilde a^2)^{1/2}
  \end{equation} 
was used, with $\tilde a=a/(1-\alpha)$.
Similar to the Wiener--Hopf solution for an isolated edge, 
at large values of the exchange parameter the density profile becomes
nonmonotonic.
Note that our density functional, being quadratic in $n(r)$,
obeys an exact particle--hole symmetry. Hence the density
profiles on the upper and lower sides of the plateau in Fig.2
are identical up to a sign change.

In conclusion, we find that finite density of localized states and 
electron exchange interaction both have similar effect on the width 
of the incompressible strip. The strip width increases
as a function of the localized states density, and as a function of 
electron exchange parameter. 
However, the density profile in these two problems evolves differently. 
For a high density of localized states the density gradient becomes 
nearly uniform, 
whereas at large exchange the plateau in the density distribution becomes
wider. At very large exchange, the density profile becomes nonmonotonic,
indicating formation of a one dimensional Wigner crystal at the edge.

L.L. is grateful to R. Ashoori, G. Finkelstein, 
T. D. Fulton, and A. Yacoby for 
useful discussions of their data. Research at MIT 
is supported in part by the MRSEC Program of NSF under 
award 6743000 IRG.

\centerline{\bf References} 
 
\noindent 
1. D. B. Chklovskii, B. I. Shklovskii, and L. I. Glazman, Phys. Rev. B 
{\bf  46}, 4026 (1992). 
 
\noindent 
2. S. H. Tessmer, P. I. Glicofridis, R. C. Ashoori, L. S. Levitov, 
M. R. Melloch, Nature, {\bf 392}, No.6671, 51 (1998);\\ 
G. Finkelstein,  P. I. Glicofridis, S. H. Tessmer, R. C. Ashoori, M. R. Melloch,
preprint 
 
\noindent 
3. A. Yacoby, H. F. Hess, T.A. Fulton, L. N. Pfeiffer and K. W. West,
Solid State Communications, {\bf 111}, 1-13 (1999) 
 
\noindent
4. K. L. McCormick, M. T. Woodside, M. Huang, M. S. Wu, P. L. McEuen, 
C. Duruoz, J. S. Harris, Phys. Rev. B, {\bf 59}, 4654 (1999) 
 
\noindent
5.  Y. Y. Wei, J. Weis, K. v. Klitzing, K. Eberl, 
Phys. Rev. Lett., {\bf 81}, 1674 (1999)
 
\noindent 
6. I. A. Larkin, J. H. Davies, Phys. Rev. B, {\bf 52}, R5535 (1995) 
 
\noindent 
7. J. P. Eisenstein, L. N. Pfeiffer, and K. W. West, Phys. Rev. 
  Lett. {\bf 68}, 674 (1992);\\ 
B. Tanatar, D. M. Ceperley, Phys. Rev. B {\bf 39}, 5005 (1989) 
 
\noindent 
8. A. L. Efros, cond-mat/9905368

\noindent 
9. I. A. Larkin, L. S. Levitov, to be published


 
 

 
 
 
 
\end{document}